\documentclass{aastex61}

\begin{document}

   \title{Characterizing microlensing planetary system OGLE-2014-BLG-0676Lb with adaptive optics imaging}\footnote{This paper includes data gathered with the 6.5m Magellan Clay Telescope at Las Campanas Observatory, Chile.}

\correspondingauthor{Subo Dong}
\email{dongsubo@pku.edu.cn}

\author{Xiao-Jia Xie}
\affiliation{Kavli Institute for Astronomy and Astrophysics, Peking University, Yi He Yuan Road 5, Hai Dian District, Beijing 100871, People's Republic of China}
\affiliation{Department of Astronomy, Peking University, Yi He Yuan Road 5, Hai Dian District, Beijing 100871, People's Republic of China}
\author{Subo Dong}
\affiliation{Kavli Institute for Astronomy and Astrophysics, Peking University, Yi He Yuan Road 5, Hai Dian District, Beijing 100871, People's Republic of China}
\author{Yossi Shvartzvald}
\affiliation{Department of Particle Physics and Astrophysics, Weizmann Institute of Science, Rehovot 76100, Israel}
\author{Andrew Gould}
\affiliation{Department of Astronomy, Ohio State University, 140 W. 18th Avenue, Columbus, OH 43210, USA}
\affiliation{Max-Planck-Institute for Astronomy, K\"{o}igstuhl 17, D-69117 Heidelberg, Germany}
\author{Andrzej Udalski}
\affiliation{Astronomical Observatory, University of Warsaw, Al. Ujazdowskie 4, 00-478 Warszawa, Poland}
\author{Jean-Philippe Beaulieu}
\affiliation{School of Natural Sciences, University of Tasmania, Private Bag 37 Hobart, Tasmania 7001 Australia}
\affiliation{Sorbonne Universit\'e, UPMC Universit\'e Paris 6 et CNRS, UMR 7095, Institut d'Astrophysique de Paris, 98 bis Bd Arago, F-75014 Paris, France}
\author{Charles Beichman}
\affiliation{Jet Propulsion Laboratory, California Institute of Technology, 4800 Oak Grove Drive, Pasadena, CA 91109, USA}
\affiliation{Caltech/IPAC-NASA Exoplanet Science Institute, 770 S. Wilson Ave, Pasadena, CA 91106, USA}
\author{Laird Miller Close}
\affiliation{Steward Observatory, University of Arizona, Tucson, AZ 85721, USA}
\author{Calen B. Henderson}
\affiliation{IPAC, Mail Code 100-22, Caltech, 1200 East California Boulevard, Pasadena, CA 91125, USA}
\author{Jared R. Males}
\affiliation{Steward Observatory, University of Arizona, Tucson, AZ 85721, USA}
\author{Jean-Baptiste Marquette}
\altaffiliation{(Associated to) Sorbonne Universit\'e, UPMC Universit\'e Paris 6 et CNRS, UMR 7095, Institut d'Astrophysique de Paris, 98 bis Bd Arago, F-75014 Paris, France}
\affiliation{Laboratoire d'Astrophysique de Bordeaux, Univ. Bordeaux, CNRS, B18N, all\'ee Geoffroy Saint-Hilaire, F-33615 Pessac, France}
\author{Katie M. Morzinski}
\affiliation{Steward Observatory, University of Arizona, Tucson, AZ 85721, USA}
\author{Christopher R. Gelino}
\affiliation{IPAC, Mail Code 100-22, Caltech, 1200 East California Boulevard, Pasadena, CA 91125, USA}
\begin{abstract}
{We constrain the host-star flux of the microlensing planet OGLE-2014-BLG-0676Lb using adaptive optics (AO) images taken by the Magellan and Keck telescopes. We measure the flux of the light blended with the microlensed source to be  $K = 16.79 \pm 0.04$\,mag and $J = 17.76 \pm 0.03$\,mag. Assuming that the blend is the lens star, we find that the host is a $0.73_{-0.29}^{+0.14}\,M_{\odot}$ star at a distance of $2.67_{-1.41}^{+0.77}$\,kpc, where the relatively large uncertainty in angular Einstein radius measurement is the major source of uncertainty. With mass of $M_p = 3.68_{-1.44}^{+0.69}\,M_J$, the planet is likely a ``super Jupiter'' at a projected separation of $r_{\perp} = 4.53_{-2.50}^{+1.49}$\,AU, and a degenerate model yields a similar $M_p = 3.73_{-1.47}^{+0.73}\,M_J$ at a closer separation of $r_{\perp} = 2.56_{-1.41}^{+0.84}$ AU.  Our estimates are consistent with the previous Bayesian analysis based on a Galactic model. OGLE-2014-BLG-0676Lb belongs to a sample of planets discovered in a ``second-generation'' planetary microlensing survey, and we attempt to systematically constrain host properties of this sample with high-resolution imaging to study the distribution of planets.}
\end{abstract}


%
%
\section{Introduction}           

Gravitational microlensing is most sensitive to discovering planets beyond the snow line \citep{Mao:1991aa,Gould:1992aa}, where giant planets are predicted to form efficiently according to the core-accretion theory \citep{Lissauer87, Pollack96}. Planet frequencies derived from microlensing planet samples have offered important insights into understanding planet formation \citep{ZhuDong:2021araa}. \citet{Shvartzvald:2016aa} published the first sample from a ``second-generation'' microlensing experiment, which only considered high-cadence data from wide-field surveys of Optical Gravitational Lensing Experiment (OGLE), Microlensing Observations in Astrophysics (MOA) and Wise to search for planets without follow-up observations. Being free from possible follow-up selection biases,  such a homogeneous sample is especially suitable for inferring the distributions of planetary systems.

The planet-to-star mass ratio $q$ is usually precisely measured from the microlensing light curves, but the lens (host) mass and distance are generally not separately constrained, posing a major limitation to studying the distributions of planets and their hosts. The lens mass and distance are related to an observable, the angular Einstein radius $\theta_\mathrm{E}$, 

\begin{equation}
\label{constraint_theta_E}
\theta_\mathrm{E}=\sqrt{\kappa M_L \pi_{\mathrm{rel}}},\qquad  \pi_{\mathrm{rel}} = {\frac{\mathrm{AU}}{D_L} - \frac{\mathrm{AU}}{D_S}},
\end{equation}
 where $M_L$ is the mass of the lens, $\pi_{\mathrm{rel}}$ is the relative parallax between the lens at distance $D_L$ and the source at $D_S$, and $\kappa = {4G}/({c^2\mathrm{AU}})=8.14\,\mathrm{mas}\,M^{-1}_\odot$ is a constant. $\theta_\mathrm{E}$ is frequently constrained in planetary microlensing events thanks to the finite source effects \citep{Gould94}, which constrain the ratio $\rho$ between angular radius of the source $\theta_*$ and $\theta_\mathrm{E}$, and $\theta_*$ can regularly be measured well from source color and flux \citep{Yoo04}. For almost all microlensing events, the Einstein crossing timescale $t_\mathrm{E} = \theta_\mathrm{E}/\mu_{\rm rel}$, where $\mu_{\rm rel}$ is the relative lens-source proper motion, is always measured well. However, with only $t_\mathrm{E}$ and $\theta_\mathrm{E}$ known, it is insufficient to break the degeneracy and directly measure lens mass/distance. It is a common practice to employ Bayesian analysis to incorporate priors of mass, distance and velocity distributions of stars from an assumed Galactic model, but doing so introduces an untested assumption that planet hosts follow the same distributions. In certain individual cases, the directly measured lens parameters differ significantly from those derived from Bayesian analysis \citep[see, e.g.,][]{Vandorou:2020}. It is therefore important to lift the degeneracy and directly measure  the lens physical parameters.

One way to break the degeneracy between lens mass and distance is measuring the lens flux, which establishes another constraint between lens mass and distance in addition to that from $\theta_\mathrm{E}$. The bulge microlensing fields are so crowded that only high-resolution imaging observations such as with the {\it Hubble Space Telescope} ({\it HST}) \citep[e.g.,][]{Dong:2009aa} and ground-based adaptive optics (AO) \citep[e.g.,][]{Janczak10} can disentangle the microlens from nearby blended stars.

Primarily using the $6.5$-m Magellan AO system (MagAO; \citealt{close2012,males2014,Morzinski2014}), we systematically observed the \citet{Shvartzvald:2016aa} planetary microlensing sample with high-resolution imaging to constrain the physical properties of the hosts and planets. Here we report our AO observations from MagAO and Keck of OGLE-2014-BLG-0676Lb from the \citet{Shvartzvald:2016aa} sample, and the detailed analysis of its light curves was reported in \citet{Rattenbury:2017aa} (hereafter R2017).


\section{Observations and Data Reduction}
\label{MagAO}

MagAO is mounted on the $6.5$-m Magellan Clay telescope at Las Campanas Observatory in Chile, which became fully operational in 2012. 
 We observed the OGLE-2014-BLG-0676 (equatorial coordinates: $\alpha=17^{\rm h}52^{\rm m}24\fs50$, $\delta=-30\degr32\arcmin54\farcs2$; Galactic coordinates: $l=359.37608\deg$, $b=-2.09306\deg$) field with MagAO's near-infrared (NIR) imager Clio2 on the night of UT 2015 May 19, and we used the wide camera, which has a pixel scale of $27.5$\,mas and a field of view (FOV) of about $28\arcsec\times14\arcsec$ \citep{Morzinski:2015aa}. The AO guide star is located at $\alpha=17^{\rm h}52^{\rm m}22\fs69$, $\delta=-30\degr32\arcmin48\farcs02$ with $K = 11.30$\,mag. We employed an octagon dithering pattern with a step size of $\sim0.5\arcsec$, and five frames were taken at each of the eight dithering positions. The detector readout time was 280 ms, and each individual science frame took 30 seconds of integration time.

\begin{figure}[h]
\plotone{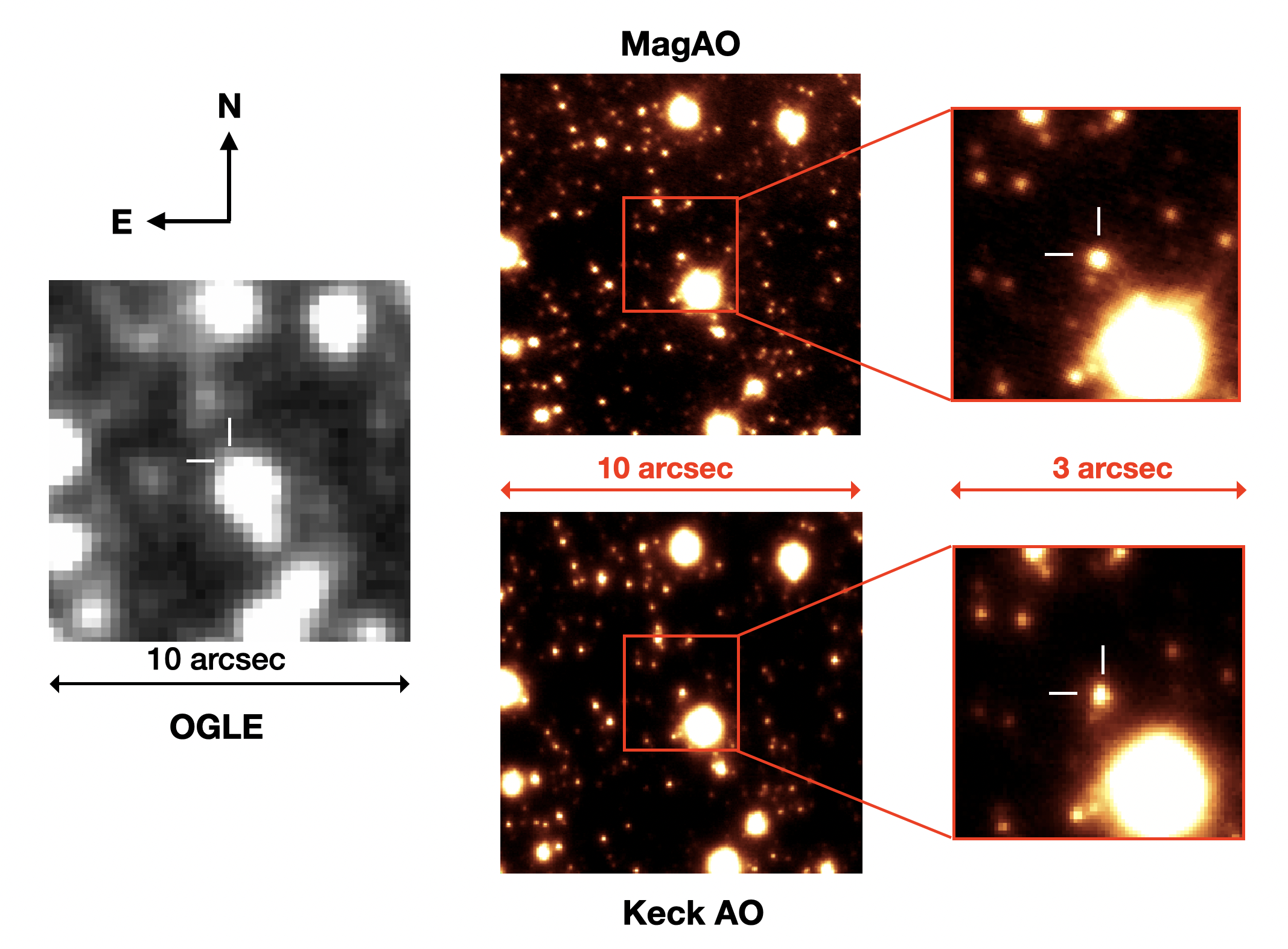}
\caption{The images taken by OGLE (left), MagAO (top right) and Keck (bottom right). The right-most panels display the zoomed MagAO and Keck images centered on the microlens, marked by {\it white half crosses}).}
\label{fig:MagAO_image}
\end{figure}

We processed the images by including corrections for nonlinearity, dark current, and flat field. Following the recipe in \cite{Morzinski:2015aa}, we made non-linearity corrections for pixels with analog to digital unites (ADUs) above $27000$. We note that none of the stars that we relied on for photometric or astrometric measurements landed on pixels exceeding this threshold. We find that the values of dark currents gradually vary as a function of time during the night. The dark and flat calibration frames were taken about three hours after the science frames of OGLE-2014-BLG-0676, and we applied constant ADU offsets to each science frame to account for the dark current changes. By examining the flat field, we find significant variations along the $x$-axis but there is no obvious trend in $y$.  For each dark-subtracted science frame, we calculate the background ADUs as a function of $x$-axis by taking medians, and we then match the trend in $x$ with that from the dark-subtracted flat with linear fitting to obtain the offsets. After making the dark and flat corrections, we performed astrometric alignment of the frames with the positions of 11 bright isolated stars, and then they were median-combined (see the upper right panel of Fig.~\ref{fig:MagAO_image}).

We observed OGLE-2014-BLG-0676 in $J$-band utilizing the NIRC2 camera and AO system on the Keck \uppercase\expandafter{\romannumeral2} telescope on UT 2016 August 12 with the wide camera with a plate scale of $39.7$\,mas. We took six dithered frames and at each dithering position there were six images with an integration time of 15 s for each image. We corrected the dark and flat field following standard procedures. Then the images were astrometrically aligned and stacked  (see the lower right  panel of Fig.~\ref{fig:MagAO_image}).

The full widths at half maximum (FWHMs) of isolated stars on the $K$-band MagAO image and $J$-band Keck image are  $150\,\mathrm{mas}$ and $130\,\mathrm{mas}$, respectively. The lens-source relative proper motion is $\mu_{\rm rel}\approx4\,\mathrm{mas\,yr^{-1}}$. At the time of MagAO and Keck observations ($1.05$\,yr and $2.29$\,yr after the peak, respectively), the lens and source were separated by about angular distances of $5-10$\,mas, which were much smaller than the FWHMs. 

\section{blended light}
\label{blend_light}

We follow the main steps laid out in \citet[][]{Janczak10} to perform astrometric and photometric analysis of the AO images. Astrometry and aperture photometry are done with SExtractor \citep{Bertin:1996aa}. Positions of eight isolated common stars are referenced for coordinate transformations between OGLE and MagAO frames. The transformed position of the source on the subtracted OGLE images is well-matched ($8\pm10$\,mas) with that of an isolated point source on the MagAO image, and the nearest neighbor is $\sim230$ mas away. Therefore we identify this point source as the microlens baseline object (marked with a white half cross in the upper right panel of Fig.~\ref{fig:MagAO_image}). We adopt an aperture size of 1.5 FWHM for photometry on each of the dithered MagAO images and 2 FWHM on the Keck images. Due to the lack of overlapping stars between AO images and Two Micron All Sky Survey (2MASS), we rely on images from the VISTA Variables in the Via Lactea (VVV) survey \citep{Minniti:2010aa} to bridge between 2MASS and AO images for performing the photometric calibration. We perform  point spread function (PSF) photometry on VVV images using DoPhot \citep{Schechter:1993aa}, and we also perform another independent set of VVV photometry according to the procedures of \citet{Beaulieu:2016aa} and find consistent results. We note that multiple $K$-band VVV images exist. These images exhibit a broad range of FWHMs, and as a result, the non-linearity thresholds and detection limits vary significantly from image to image. We perform internal calibration of $K$-band VVV images to place them into the same instrumental system and then combine them to generate a single photometric catalog.
In comparison, the $J$-band VVV images have similar FWHMs and thus do not have such an issue. To calibrate the VVV magnitudes into the standard system, we rely on 2MASS comparison stars fainter than $K = 11.3$\,mag and $J = 13.0$\,mag to avoid the non-linearity effect on VVV detectors. Then we calibrate the MagAO instrumental magnitude utilizing bright and isolated common stars between MagAO and VVV images. To evaluate how isolated a VVV star is, we identify nearby MagAO stars and calculate the flux ratio between the star of interest and the sum of all nearby contributors. In addition, within $\sim 1\arcsec$ of a VVV star, if there is any MagAO star within that is bright but unidentified by DoPhot on the VVV image, we exclude it from our selection of comparison stars. Finally,  we choose six common bright and isolated stars to calibrate the MagAO instrumental magnitude with VVV. We perform similar calibration on Keck instrumental magnitude, and due to the much larger FOV of Keck, a larger number of 98 common stars are available. The calibrated magnitude of the microlens baseline object is $K_{\mathrm{base}}=16.72\pm0.04$ and $J_{\mathrm{base}}=17.72\pm0.03$ in the 2MASS magnitude system. 

\begin{figure*}[h]
\plotone{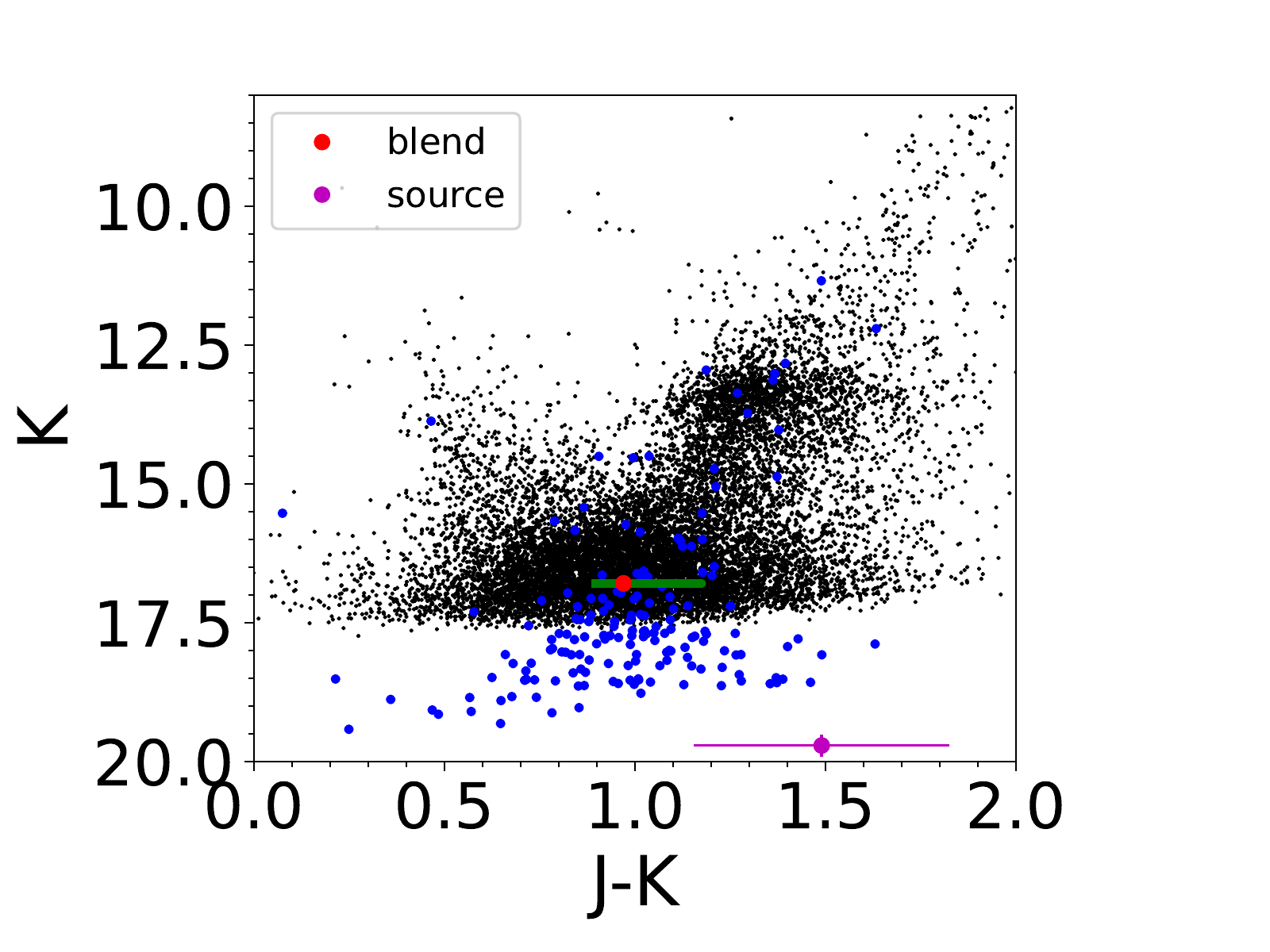}
\caption{The $J-K$ vs. $K$ color-magnitude diagram of the OGLE-2014-BLG-0676 field. Most of the stars (marked as {\it black dots}) are within $3\arcmin$ of the microlens position from VVV. For stars with  $K<11.3$\,mag, they either saturate or reach non-linearity on VVV, and 2MASS magnitudes are displayed. The stars detected on both MagAO ($K-$band) and Keck ($J-$band) images are represented as {\it blue dots}. The blend and source are signified as {\it red} and {\it magenta dots} with error bars, respectively. The {\it green line} indicates the range of $J-K$ color expected for the lens + companion scenario discussed in Sect.~\ref{sec:lenscom}.}
\label{fig:CMD}
\end{figure*}

R2017 found that the source is faint ($I_s = 23.5\pm0.1$\,mag), and hence its contribution to the microlens baseline flux is expected to be small. We estimate $K$- and $J$-band flux of the source star using its $I$-band flux and main-sequence isochrones. The distance modulus and extinction of red clumps toward the direction of the microlens are ${\rm DM} = 14.56\pm0.15$ \citep{Nataf:2013aa} and $(A_I, E(V-I)) = (2.50, 2.09)\pm (0.05, 0.07)$, respectively, according to R2017. Assuming that the source is at the same distance as the red clumps, we infer the source $I$-band absolute magnitude to be $M_{s, I}=6.4\pm0.2$, which is consistent with a late-type main-sequence star. Motivated by the metallicity and age distribution measured by \citet{Bensby:2013aa}, we adopt a range of parameters encompassing typical bulge dwarfs: $\mathrm{[Fe/H]=[-0.6,0.5]}$, $\mathrm{age}=[5,15]\,\mathrm{Gyr}$ and generate isochrones from Dartmouth Stellar Evolution Database \citep{Dotter:2008aa} to estimate the source $K$- and $J$-band absolute magnitude as $M_{s,K}=4.72\pm0.18$ and $M_{s,J}=5.50\pm0.18$ respectively, where the error mainly comes from the uncertainty in $I$-band absolute magnitude and the uncertainties in metallicity.  Toward the direction of OGLE-2014-BLG-0676, we estimate that $A_K=0.37\pm0.08$ and $A_J=1.08\pm0.20$  \citep{Nishiyama:2009aa,Gonzalez:2012aa}. Therefore, we estimate the source brightness to be $K_s=19.71\pm{0.20}$ and $J_s=21.20\pm0.27$, which is $\gtrsim 3$\,mag fainter than the microlens baseline object derived from the AO images. Subtracting the source flux from the baseline object, we find that the blend has $K_{b}=16.79\pm0.04$ and $J_{b}=17.76\pm0.03$. 

\section{Physical Constraints}
\label{lens}

In this section, we use our measured blend fluxes to place physical constraints on the system. By counting stars brighter than the $K$-band blend on the MagAO, we estimate the number density to be $0.18\,\mathrm{arcsec}^{-2}$. The chance for a random star coinciding within $10$\,mas ($1\sigma$ upper limit) of the source position is $\approx5\times10^{-5}$, so it is unlikely that the blend is unrelated to the microlensing event. The blend can be due to the lens, a lens companion or a source companion or the combination of any of these possibilities. In the following, we consider these three scenarios separately, and compare the relative probability of the latter two scenarios with respect to that of the lens in \S~\ref{sec:sourcecom} and \S~\ref{sec:lenscom}, respectively.

\subsection{Blend = Lens}
\label{blend_lens}
Assuming that the blended light is due to the lens star, the lens has $K_L=K_{b}=16.79\pm0.04$ and $J_L=J_{b}=17.76\pm0.03$. Using the isochrones, we sample the lens distance and mass on a grid  of $0.01\,\mathrm{kpc}<D_L<8\,\mathrm{kpc}$, $0.12M_\odot<M_L<1.2M_\odot$. We adopt uniform distributions of $\mathrm{[Fe/H]}$ between $-0.2$ and $+0.2$ and age between 1 and 10 Gyr. Following \citet{Bennett20}, we model extinction toward the lens $A_L$ as a function of $D_L$ by $A_L=(1-\exp{(-D_L |\sin{b}| /h_{\rm dust}}))/(1-\exp{(-D_S |\sin{b}|/h_{\rm dust}}))A_S$, where $h_{\rm dust}=0.1\,{\rm kpc}$  is the scale height of the dust and $A_S$ is the extinction toward the source.

The other constraint is from the measurement of the angular Einstein radius $\theta_\mathrm{E}$. R2017 obtained $\theta_\mathrm{E} = \theta_*/\rho = 1.38\pm0.43$\,mas, where $\rho=(2.78\pm0.33)\times10^{-4}$ based on finite-source effects and $\theta_* = 0.38\pm0.11\,\mu\mathrm{as}$ was the adopted angular source size that was derived from an estimated color of the source $(V-I)_s = 4.27\pm0.11$. The $V$-band light curve was not available to R2017, who inferred the color utilizing the broad-band ($R/I$) MOA data following the method of \citet{Gould:2010ab}. However, we find that such a star (with $(V-I)_s = 4.27\pm0.11$ and the measured source flux of $I_s = 23.5$) substantially deviates from the isochrones expected for bulge stars. Considering the isochrone analysis on the source star discussed in \S~\ref{blend_light}, we find that, for the measured source flux $I_s = 23.51\pm0.11$, the expected source color is instead $(V-I)_s = 3.53\pm0.25$ with the main source of color budget being the uncertainties in metallicity estimates. Our isochrone analysis indicates that the source radius is $\mathrm{log}(R_s/R_{\odot})=-0.21\pm0.04$ and thus the angular source radius is $0.35\pm0.04\,\mu\mathrm{as}$ using ${\rm DM}=14.56\pm0.15$. Combining $\rho=(2.78\pm0.33)\times10^{-4}$ from R2017,  we obtain $\theta_\mathrm{E} = 1.26\pm0.20$\,mas, which is consistent with the R2017 value and adopted in our subsequent analysis.

\begin{figure*}
\begin{minipage}[t]{1.0\linewidth}
\plotone{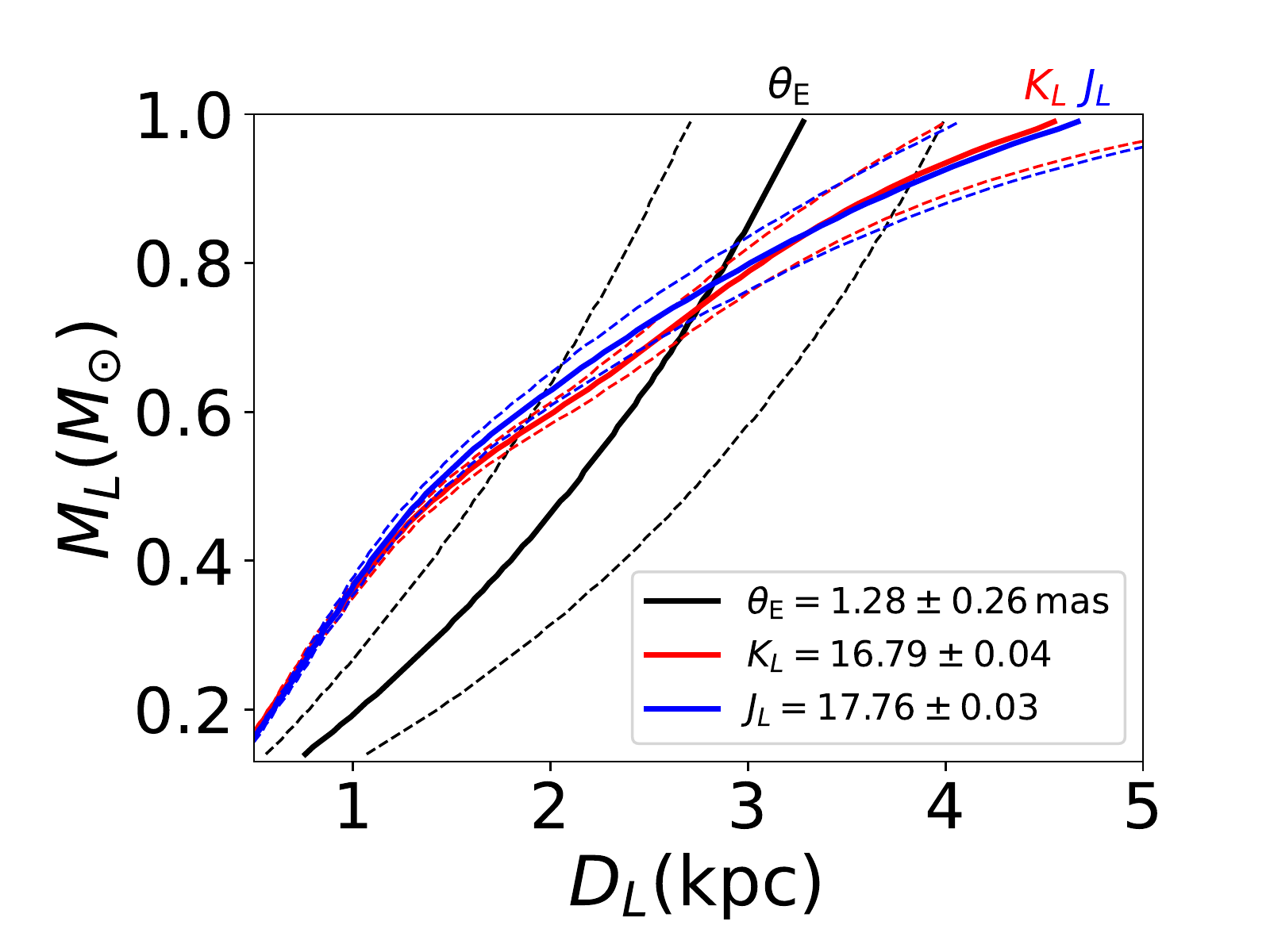}
\end{minipage}
\begin{minipage}[t]{0.49\linewidth}
\centering
\includegraphics[width=80mm]{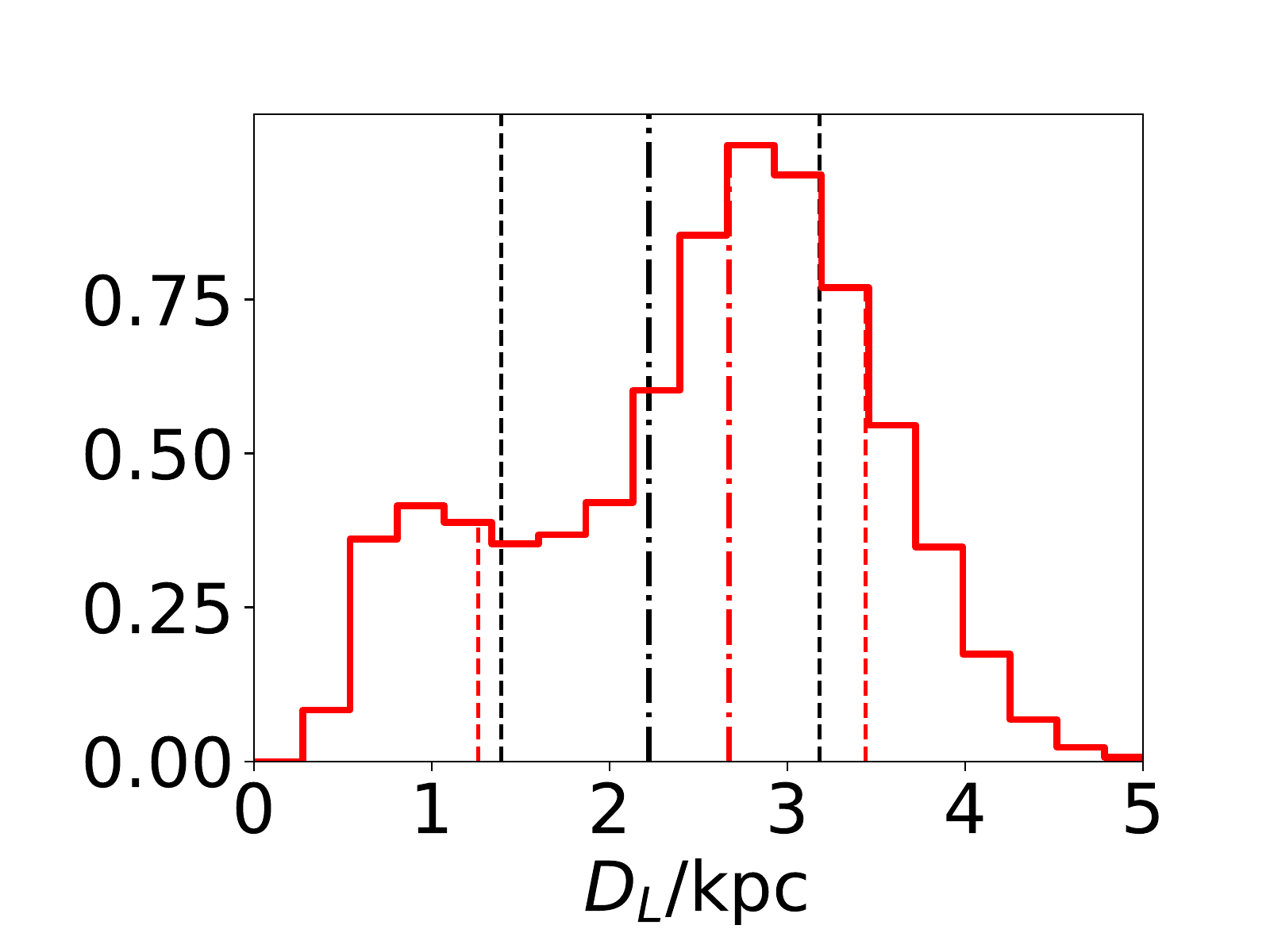}
\end{minipage}
\begin{minipage}[t]{0.49\linewidth}
\centering
\includegraphics[width=80mm]{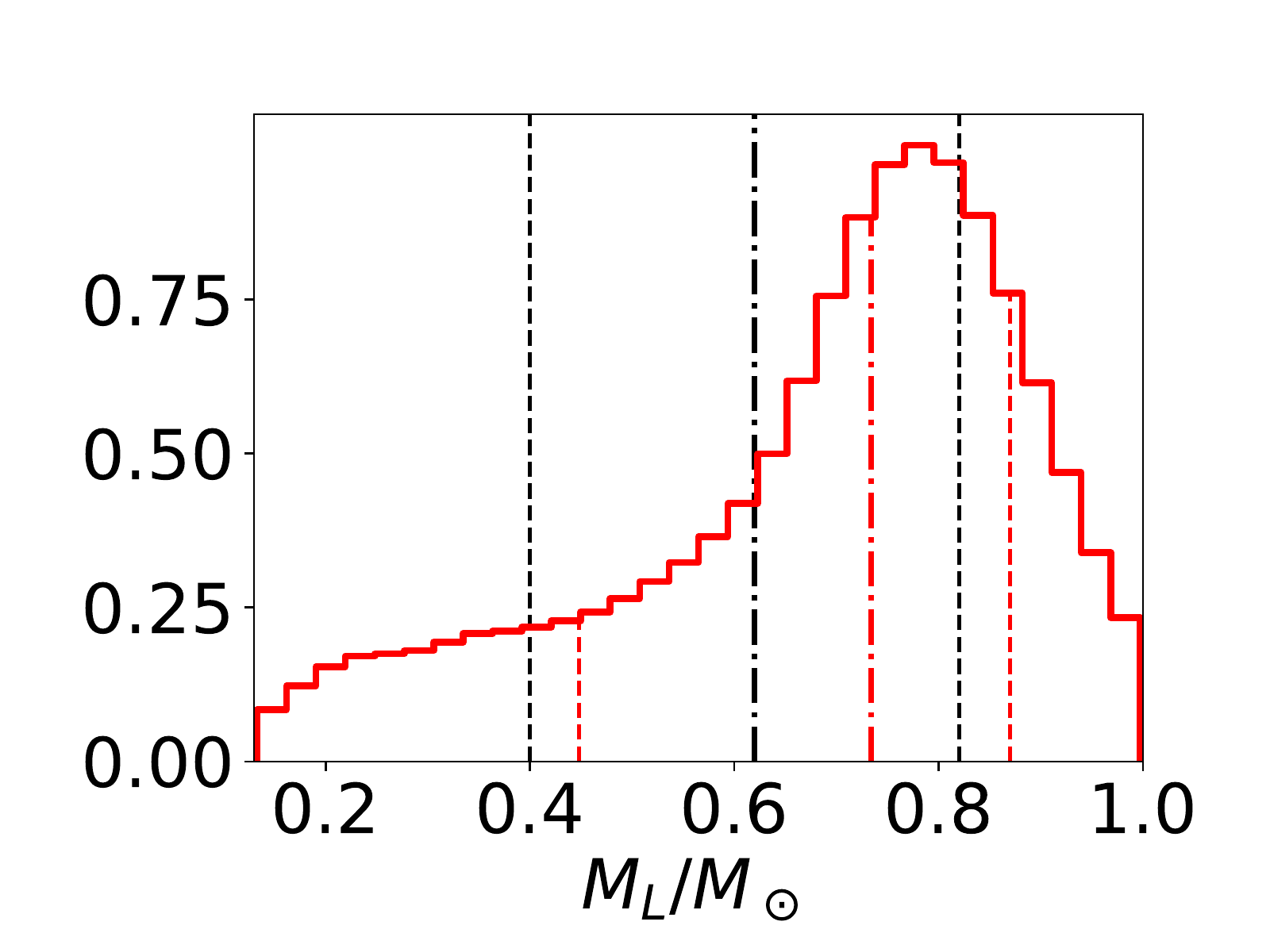}
\end{minipage}
\caption{The upper panel features the different constraints on lens mass and distance. The {\it black, red} and {\it blue lines} stand for constraints of $\theta_\mathrm{E}$, lens $K$ band flux and lens $J$-band flux, respectively. The {\it dashed lines} signify $1\sigma$ error. {\it Red histograms} in the bottom two panels depict the differential likelihood distributions of lens distance (left) and mass (right). The {\it vertical red dash-dotted} and {\it dashed lines} correspond to the median and 68\% confidence intervals of parameters, respectively. The {\it black dash-dotted} and {\it dashed lines} show the parameters inferred in R2017, constrained by $\theta_\mathrm{E}$ and upper limit of lens $I$-band flux using the Galactic model priors.}
\label{fig:likelihood}
\end{figure*}

We combine the constraints from $K$- and $J$-band lens fluxes using main sequence isochrones \citep{Dotter:2008aa} and the measured $\theta_\mathrm{E}$, which are shown in yellow, red and black, respectively, in the upper panel of Fig.~\ref{fig:likelihood}. The joint posterior probability distributions of the lens distance and mass are displayed in the bottom panel of Fig.~\ref{fig:likelihood}. We note that the $K$ and $J$-band constraints closely track each other, so that the two-band NIR observations essentially provide redundant information. We find that the lens star is an $M_L = 0.73_{-0.29}^{+0.14}\,M_{\odot}$ star at a distance $2.67_{-1.41}^{+0.77}$\,kpc. Our results are broadly consistent with those inferred by R2017 applying Bayesian analysis assuming priors based on the Galactic model: $M_L=0.62_{-0.22}^{+0.20}\,M_{\odot}$ and $D_L=2.22_{-0.83}^{+0.96}$\,kpc.  We note that the major source of uncertainty in our physical parameter determinations comes from the relatively large errorbar of $\theta_\mathrm{E} = 1.26\pm0.20$\,mas, which is due to the unusually poorly measured $\rho$ compared to most other planetary microlensing events. Assuming that the uncertainty in $\theta_\mathrm{E}$ is half of its present value, that is, if $\theta_\mathrm{E} = 1.26\pm0.10$\,mas, then we would obtain significantly more precise physical parameters of $M_L = 0.74\pm{0.09}\,M_{\odot}$ and $D_L = 2.70 \pm {0.45}$\,kpc. Therefore, for a planetary microlensing event with $\theta_\mathrm{E}$ measured at a typical accuracy, follow-up observations like ours could yield fairly precise lens mass and distance determinations.

Using the planet-to-star mass ratio ($q = 4.78\pm0.13\times10^{-3}$ for the  ``wide'' solution with $s>1$ and $q = 4.85\pm0.31\times10^{-3}$ for the degenerate  ``close'' solution with $s<1$) and the projected planet-to-star separation in the units of Einstein radius $s$ derived from R2017, we obtain the planetary mass $M_p = 3.68_{-1.44}^{+0.69}\,M_J$ and the projected planet-to-star separation $r_{\perp} = 4.53_{-2.50}^{+1.49}$\,AU for the wide solution, and $M_p = 3.73_{-1.47}^{+0.73}\,M_J$, $r_{\perp} = 2.56_{-1.41}^{+0.84}$\,AU for the close solution. 

\subsection{Source Companion}
\label{sec:sourcecom}
If the blend were a companion of the source, based on the $J-K$ vs. $K$ color-magnitude diagram (depicited in Fig.~\ref{fig:CMD}), the K-dwarf source would have a subgiant companion, and in this case the lens would be less luminous than blend. In the following, we first evaluate the former probability, and then we incorporate the latter condition to calculate the relative probability with respect to the blend being the lens.

Subgiants evolve from G dwarfs, and the time that stars stay on the subgiant and giant branches counts only 10\% compared to that of the main sequence. We subsequently utilize the multiplicity statistics for local FGK dwarfs \citep{Raghavan2010} to approximately estimate the probability for such a binary.  Based on mass-ratio distribution, roughly $20\%$ of G dwarfs have K dwarf companions.  Our astrometric measurement suggests that the source companion needs to lie within $\approx 30$\,mas ($\approx 250$ AU) of the source star, and in order to avoid detectable effects from a binary-source event, the binary separation is required to be $\gtrsim 10$\,AU (i.e., greater than $\sim2$ Einstein radii). Such a separation range encompasses about $\sim 50\%$ of binaries. Therefore, the probability that the source hosts a companion with flux and separation consistent with our measurements is $\sim 1\%$. 

In this scenario, the lens needs to be less luminous than the blend while still subjects to the constraint from the measured microlensing event parameters, so its mass needs to be less than $0.44 M_{\odot}$ at $1\sigma$. Then we estimate the relative probability of this condition with the condition that the lens mass is between $0.44 M_{\odot}$ and $0.87 M_{\odot}$ considering the Bayesian analysis results by R2017 and found 0.29. Therefore, the blend being the lens is preferred over being the source companion by a factor of $\sim 300$.

\subsection{Lens Companion}
\label{sec:lenscom}
In this section, we discuss the scenario that the blend is the combination of the lens and a companion and estimate the relative probability with the scenario that the blend is dominated by the lens.

 The absence of additional binary-lens signals (besides the reported planetary signal) sets a lower limit on the lens separation. OGLE-2014-BLG-0676 is a well-covered high-magnification event with peak single-lens magnification of $A_{\rm max} \approx 250$. For a distant companion, it would induce shear perturbation scaled with ${q}/{s^2}$ \citep{Chang:1979aa,Chang:1984aa} at the peak of a high-magnification event. \citet{Janczak10} carried out triple-lens simulations to calculate the shear that could be ruled out using a threshold of $\Delta{\chi^2} = 9$. We make a more simplified calculation by relying on binary-lens simulations without the presence of the detected planet. Given our simplification, we use a more stringent threshold of  $\Delta{\chi^2} = 25$. Employing the ``mapmaking'' algorithm \citep{Dong:2006, Dong:2009bb}, we calculate the detection efficiency utilizing the light curve of OGLE-2014-BLG-0676 in R2017 and evaluate a range of $q$ from $10^{-1}$ to $1$. We find that systems with $q/s^2>10^{-2.9}$ can be securely ruled out. The astrometric constraint translates into an upper limit of the separation between the light centroid of the lens systems (i.e., the host and its companion) and the host in the units of Einstein radius of $\lesssim 23$.  Furthermore, for a given distance, the combined light of the lens and its companion is constrained by the blend flux measurement, and at the same time, the lens mass is still subject to the $\theta_{\rm E}$ constraint. We find that a main-sequence nearly-equal-mass binary with primary mass between $\sim 0.2\,M_{\odot}$ and $\sim 0.4\,M_{\odot}$ located at $\sim 0.7 \rm{-} 2.0$\,kpc can satisfy these constraints. Combining with the shear and astrometric constraints discussed above, the binary separation needs to be in a relatively narrow range between $\sim 45$ and $\sim 80$\,AU. We estimate that the expected NIR color of such a binary is in the range of $J-K \approx  0.9 \rm{-} 1.2$ (represented as the green line in Fig.~\ref{fig:CMD}), which is broadly consistent with the measured color of the blend by MagAO and Keck. Therefore, we cannot definitively rule out the possibility that the blend is composed of a lens and a binary companion. The multiplicity of M dwarf stars is about 30\% \citep{Duchene:2013aa}, and among such binaries,  only a few percent of them are within the required separation range of $\sim 45 \rm{-} 80\,AU$.
 
In comparison, if the blend is dominated by the lens itself, it has a $60\%$ chance of being a single star according to  \citet{Raghavan2010}, while the relative probability compared to a $0.2-0.4\,M_\odot$ lens is about 5 times higher according to the Galactic model, therefore the relative probability of the blend being the lens is about two orders of magnitude higher than that of the lens with a companion scenario.

\section{Discussion}
\label{discussion}

Future high-resolution imaging follow ups can determine the proper motion of the blend. However, only the lens-source relative proper motion is known ($\mu_{\rm rel} =  3.95 \pm 0.75\,{\rm mas\,yr^{-1}}$ for the wide solution or $\mu_{\rm rel} =  4.29 \pm 0.82\,{\rm mas\,yr^{-1}}$ for the close solution), and the proper motion of the faint source is not known. Unless the source is separately resolved in the future with multiple-epoch observations to determine its proper motion, no definitive test can be made considering the blend proper motion alone. When the lens and source can be separately resolved, this will yield a direct measurement of $\mu_{\rm rel}$ and thus a more accurate determination of $\theta_E = \mu_{\rm rel}*t_{\rm E}$ than currently available, which will enable a better mass estimate of the lens. At that time, whether the source hosts a companion can be directly observed, and spectroscopic observations \citep[e.g.,][]{Han19} can be used to distinguish whether the lens system is single or binary.

We identify a faint star $\sim0.23\arcsec$ southeast to the baseline object on the MagAO and Keck images. We measure its flux on the combined MagAO as $K\sim18$\,mag. Here we briefly examine the possibility that this is a binary companion of the lens. On the lens plane, its separation translates into $\sim 180\,\theta_{\rm E}$. Its flux suggests  a $\sim 0.5\,M_\odot$ M dwarf, so the shear it could induce is more than 2 orders of magnitude too small to be detected. Thus we cannot rule out the possibility that the planetary system may have a wide binary companion. Future high resolution images can be used to check if this star is indeed bound to the lens (or source) by measuring their proper motions.

As an event in the complete planetary microlensing sample of \citet{Shvartzvald:2016aa}, the measurement of the mass and distance of the OGLE-2014-BLG-0676Lb system can contribute to the statistical study of microlensing planets. In particular, our mass estimate ($M_p = 3.68_{-1.44}^{+0.69}\,M_J$) of OGLE-2014-BLG-0676Lb indicates that it probably belongs to the super-Jupiter population (while it is more massive than a Jupiter mass only at the $1.9 \sigma$ level), which was a focus of the statistical study of \citet{Shvartzvald:2016aa}. \citet{Dong:2009aa} found that the second microlensing planet OGLE-2005-BLG-071Lb \citep{Udalski:2005} is a super Jupiter around an M dwarf, and it poses a challenge to the core-accretion model of planet formation, which predicts that massive planets cannot efficiently form around M dwarfs \citep[e.g.,][]{Laughlin2004}. The mass measurement by \citet{Dong:2009aa} has been recently confirmed by \citet{Bennett20} using Keck AO follow-ups. Several more microlensing super Jupiters around likely M dwarf hosts have been discovered since OGLE-2005-BLG-071Lb. According to the recent study by \citet{Ryu:2021}, some binary-lens short-duration events may be  brown dwarfs hosting planets that are sub Jupiters and super Jupiters. \citet{Shvartzvald:2016aa} suggested a possible frequency deficit in the super-Jupiter population at $q\sim10^{-2}$ based on the inferred mass-ratio function, which corresponds to the range of $3\,M_J<M_p<13\,M_J$ assuming a typical primary mass of $0.3M_{\odot}$. While the mass ratio of OGLE-2014-BLG-0676Lb $q \approx 4.8\times10^{-3}$ is lower than the deficit identified by \citet{Shvartzvald:2016aa} , its physical mass $M_p = 3.68_{-1.44}^{+0.69}\,M_J$, is at the lower part of the above-mentioned range. However, its best-fit primary mass $M_L = 0.73\,M_{\odot}$ corresponds to a K dwarf rather than an M dwarf at $0.3M_{\odot}$ assumed by \citet{Shvartzvald:2016aa}. Once the physical properties of the hosts in the \cite{Shvartzvald:2016aa} sample are measured without making use of any priors in the Galactic model, we can verify the possible existence of the super-Jupiter deficiency and analyze planet frequency as a function of host mass and distance.

\begin{acknowledgements}
We acknowledge the support by National Key R\&D Program of China (No. 2019YFA0405100),  the China Manned Space Project with NO. CMS-CSST-2021-A11 and Project 11573003 supported by  the National Natural Science Foundation of China (NSFC). This research uses data obtained through the Telescope Access Program (TAP). This work was supported by a NASA Keck PI Data Award, administered by the NASA Exoplanet Science Institute. Data presented herein were obtained at the W. M. Keck Observatory from telescope time allocated to the National Aeronautics and Space Administration through the agency’s scientific partnership with the California Institute of Technology and the University of California. The Observatory was made possible by the generous financial support of the W.M. Keck Foundation. The authors wish to recognize and acknowledge the very significant cultural role and reverence that the summit of Maunakea has always had within the indigenous Hawaiian community. We are most fortunate to have the opportunity to conduct observations from this mountain. The OGLE project has received funding from the National Science Centre, Poland, grant MAESTRO 2014/14/A/ST9/00121 to AU. JPB and JBM acknowledge the financial support of the ANR COLD WORLDS (ANR-18-CE31-0002).  KMM's work is supported by the NASA Exoplanets Research Program (XRP) by cooperative agreement NNX16AD44G.  JPB is supported by the University of Tasmania through the UTAS Foundation and the endowed Warren Chair in Astronomy.

\end{acknowledgements}

\bibliographystyle{raa}
\bibliography{ms2021-0166}

\begin{thebibliography}{37}
\providecommand\natexlab[1]{#1}
\providecommand\JournalTitle[1]{#1}

\bibitem[{Beaulieu} {et~al.}(2016)]{Beaulieu:2016aa}
{Beaulieu}, J.-P., {Bennett}, D.~P., {Batista}, V., {et~al.} 2016, \apj, 824,
  83

\bibitem[{Bennett} {et~al.}(2020)]{Bennett20}
{Bennett}, D.~P., {Bhattacharya}, A., {Beaulieu}, J.-P., {et~al.} 2020, \aj,
  159, 68

\bibitem[{Bensby} {et~al.}(2013)]{Bensby:2013aa}
{Bensby}, T., {Yee}, J.~C., {Feltzing}, S., {et~al.} 2013, \aap, 549, A147

\bibitem[{Bertin} \& {Arnouts}(1996)]{Bertin:1996aa}
{Bertin}, E., \& {Arnouts}, S. 1996, \aaps, 117, 393

\bibitem[{Chang}(1984)]{Chang:1984aa}
{Chang}, K. 1984, \aap, 130, 157

\bibitem[{Chang} \& {Refsdal}(1979)]{Chang:1979aa}
{Chang}, K., \& {Refsdal}, S. 1979, \nat, 282, 561

\bibitem[{Close} {et~al.}(2012)]{close2012}
{Close}, L.~M., {Males}, J.~R., {Kopon}, D.~A., {et~al.} 2012, in \procspie,
  Vol. 8447, Adaptive Optics Systems III, 84470X

\bibitem[{Dong} {et~al.}(2006)]{Dong:2006}
{Dong}, S., {DePoy}, D.~L., {Gaudi}, B.~S., {et~al.} 2006, \apj, 642, 842

\bibitem[{Dong} {et~al.}(2009{\natexlab{a}})]{Dong:2009bb}
{Dong}, S., {Bond}, I.~A., {Gould}, A., {et~al.} 2009{\natexlab{a}}, \apj, 698,
  1826

\bibitem[{Dong} {et~al.}(2009{\natexlab{b}})]{Dong:2009aa}
{Dong}, S., {Gould}, A., {Udalski}, A., {et~al.} 2009{\natexlab{b}}, \apj, 695,
  970

\bibitem[{Dotter} {et~al.}(2008)]{Dotter:2008aa}
{Dotter}, A., {Chaboyer}, B., {Jevremovi{\'c}}, D., {et~al.} 2008, \apjs, 178,
  89

\bibitem[{Duch{\^e}ne} \& {Kraus}(2013)]{Duchene:2013aa}
{Duch{\^e}ne}, G., \& {Kraus}, A. 2013, \araa, 51, 269

\bibitem[{Gonzalez} {et~al.}(2012)]{Gonzalez:2012aa}
{Gonzalez}, O.~A., {Rejkuba}, M., {Zoccali}, M., {et~al.} 2012, \aap, 543, A13

\bibitem[{Gould}(1994)]{Gould94}
{Gould}, A. 1994, \apjl, 421, L71

\bibitem[{Gould} {et~al.}(2010)]{Gould:2010ab}
{Gould}, A., {Dong}, S., {Bennett}, D.~P., {et~al.} 2010, \apj, 710, 1800

\bibitem[{Gould} \& {Loeb}(1992)]{Gould:1992aa}
{Gould}, A., \& {Loeb}, A. 1992, \apj, 396, 104

\bibitem[{Han} {et~al.}(2019)]{Han19}
{Han}, C., {Yee}, J.~C., {Udalski}, A., {et~al.} 2019, \aj, 158, 102

\bibitem[{Janczak} {et~al.}(2010)]{Janczak10}
{Janczak}, J., {Fukui}, A., {Dong}, S., {et~al.} 2010, \apj, 711, 731

\bibitem[{Laughlin} {et~al.}(2004)]{Laughlin2004}
{Laughlin}, G., {Bodenheimer}, P., \& {Adams}, F.~C. 2004, \apjl, 612, L73

\bibitem[{Lissauer}(1987)]{Lissauer87}
{Lissauer}, J.~J. 1987, \icarus, 69, 249

\bibitem[{Males} {et~al.}(2014)]{males2014}
{Males}, J.~R., {Close}, L.~M., {Morzinski}, K.~M., {et~al.} 2014, \apj, 786,
  32

\bibitem[{Mao} \& {Paczynski}(1991)]{Mao:1991aa}
{Mao}, S., \& {Paczynski}, B. 1991, \apjl, 374, L37

\bibitem[{Minniti} {et~al.}(2010)]{Minniti:2010aa}
{Minniti}, D., {Lucas}, P.~W., {Emerson}, J.~P., {et~al.} 2010, \na, 15, 433

\bibitem[{Morzinski} {et~al.}(2014)]{Morzinski2014}
{Morzinski}, K.~M., {Close}, L.~M., {Males}, J.~R., {et~al.} 2014, in
  \procspie, Vol. 9148, Adaptive Optics Systems IV, 914804

\bibitem[{Morzinski} {et~al.}(2015)]{Morzinski:2015aa}
{Morzinski}, K.~M., {Males}, J.~R., {Skemer}, A.~J., {et~al.} 2015, \apj, 815,
  108

\bibitem[{Nataf} {et~al.}(2013)]{Nataf:2013aa}
{Nataf}, D.~M., {Gould}, A., {Fouqu{\'e}}, P., {et~al.} 2013, \apj, 769, 88

\bibitem[{Nishiyama} {et~al.}(2009)]{Nishiyama:2009aa}
{Nishiyama}, S., {Tamura}, M., {Hatano}, H., {et~al.} 2009, \apj, 696, 1407

\bibitem[{Pollack} {et~al.}(1996)]{Pollack96}
{Pollack}, J.~B., {Hubickyj}, O., {Bodenheimer}, P., {et~al.} 1996, \icarus,
  124, 62

\bibitem[{Raghavan} {et~al.}(2010)]{Raghavan2010}
{Raghavan}, D., {McAlister}, H.~A., {Henry}, T.~J., {et~al.} 2010, \apjs, 190,
  1

\bibitem[{Rattenbury} {et~al.}(2017)]{Rattenbury:2017aa}
{Rattenbury}, N.~J., {Bennett}, D.~P., {Sumi}, T., {et~al.} 2017, \mnras, 466,
  2710 (R2017)

\bibitem[{Ryu} {et~al.}(2021)]{Ryu:2021}
{Ryu}, Y.-H., {Hwang}, K.-H., {Gould}, A., {et~al.} 2021, arXiv e-prints,
  arXiv:2104.07906

\bibitem[{Schechter} {et~al.}(1993)]{Schechter:1993aa}
{Schechter}, P.~L., {Mateo}, M., \& {Saha}, A. 1993, \pasp, 105, 1342

\bibitem[{Shvartzvald} {et~al.}(2016)]{Shvartzvald:2016aa}
{Shvartzvald}, Y., {Maoz}, D., {Udalski}, A., {et~al.} 2016, \mnras, 457, 4089

\bibitem[{Udalski} {et~al.}(2005)]{Udalski:2005}
{Udalski}, A., {Jaroszy{\'n}ski}, M., {Paczy{\'n}ski}, B., {et~al.} 2005,
  \apjl, 628, L109

\bibitem[{Vandorou} {et~al.}(2020)]{Vandorou:2020}
{Vandorou}, A., {Bennett}, D.~P., {Beaulieu}, J.-P., {et~al.} 2020, \aj, 160,
  121

\bibitem[{Yoo} {et~al.}(2004)]{Yoo04}
{Yoo}, J., {DePoy}, D.~L., {Gal-Yam}, A., {et~al.} 2004, \apj, 603, 139

\bibitem[{Zhu} \& {Dong}(2021)]{ZhuDong:2021araa}
{Zhu}, W., \& {Dong}, S. 2021, \araa, 59, 291

\end{thebibliography}

\end{document}